\documentclass[aps,prb,twocolumn,amsmath,amssymb,superscriptaddress]{revtex4-1}

\usepackage{graphicx}
\usepackage{bm}

\pdfoutput=1 

\begin{document}

\title{Effect of external conditions on the structure of scrolled graphene edges}

\author{M. M. Fogler}
\affiliation{Department of Physics, University of California
San Diego, La Jolla, 9500 Gilman Drive, California 92093}

\author{A. H. Castro Neto}
\affiliation{Department of Physics, 590 Commonwealth Avenue,
Boston University, Boston, Massachusetts, MA 02215}

\author{F. Guinea}
\affiliation{Instituto de Ciencia de Materiales de Madrid, CSIC, Sor Juana In\'es de la Cruz 3, E28049 Madrid, Spain}

\begin{abstract}

Characteristic dimensions of carbon nanoscrolls --- ``buckyrolls'' --- are calculated by analyzing the competition between elastic, van der Waals, and electrostatic energies for representative models of suspended and substrate-deposited graphene samples. The results are consistent with both atomistic simulations and experimental observations of scrolled graphene edges. Electrostatic control of the wrapping is shown to be practically feasible and its possible device applications are indicated.

\end{abstract}

\maketitle



Carbon nanoscrolls (CNSs) are intriguing materials that differ from both carbon nanotubes and carbon nanoribbons. Unlike nanotubes, CNSs have no caps nor their electronic states are subject to periodic boundary conditions. Unlike nanoribbons, CNSs are curved, and so they respond differently to uniform external fields. CNSs have additional mechanical degrees of freedom, e.g., the inner and outer diameters whose manipulation may be utilized in many new applications.

\begin{figure}[b]
\begin{center}
\includegraphics[width=7cm]{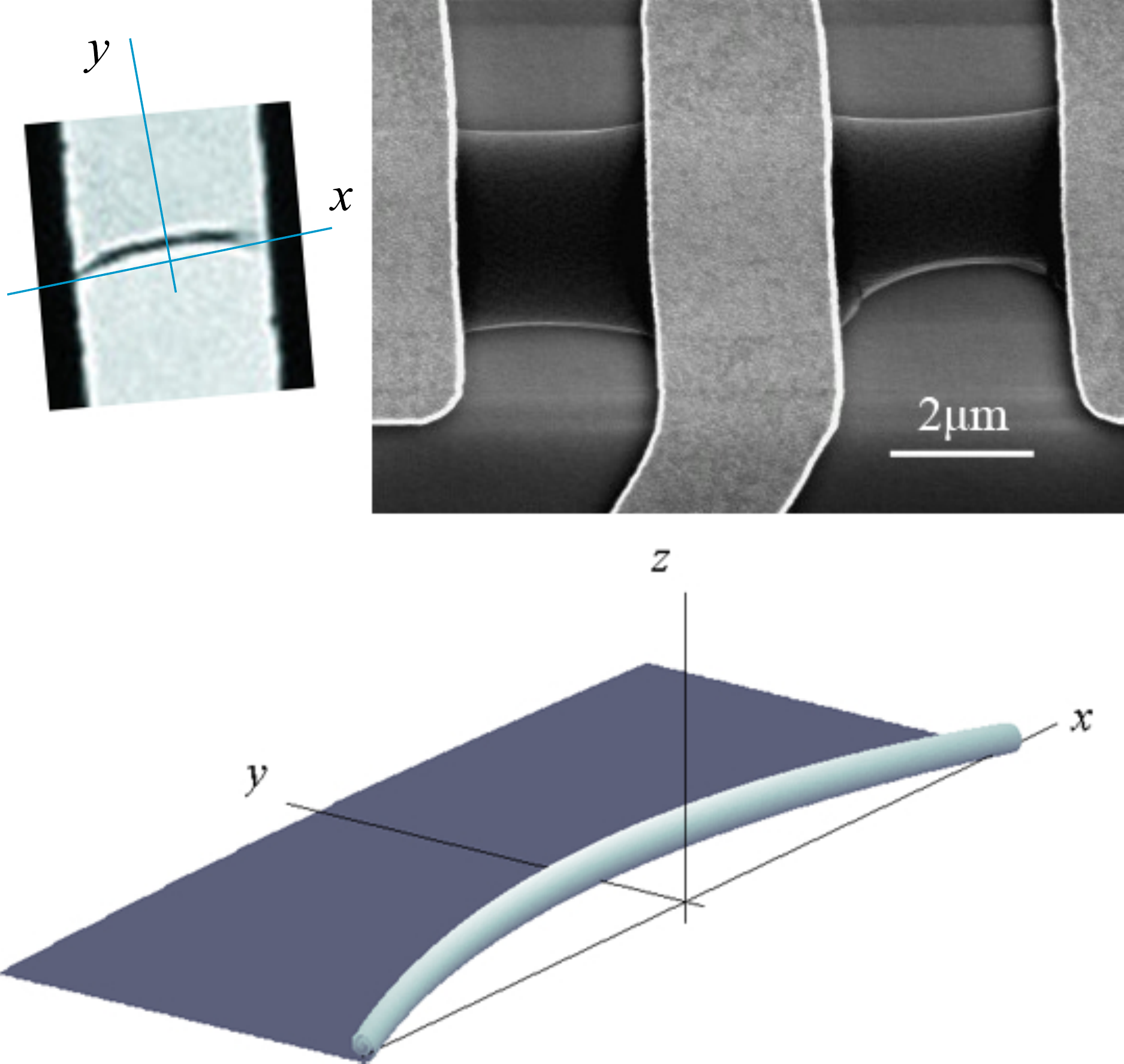}
\end{center}
\caption{(Color online) Left: a fragment of Fig.~S1 from Ref.~\onlinecite{Meyer2007tso} showing an edge CNS, on which we overlaid the $x$-$y$ axes to bring out its parabolic shape. Bottom: an illustrative sketch of the same structure. Right: another suspended graphene device with scrolled edges (courtesy of A.~K.~Geim).
\label{fig:clamped}
}
\end{figure}

Fabrication methods of CNSs include arc discharge,~\cite{Bacon1960gsa} sonication of graphite,~\cite{Viculis2003acr, Savoskin2007cnp} scratching it with an AFM tip,~\cite{Prinz2007maa} electrodeposition of graphene in a gaseous atmosphere,~\cite{Sidorov2009edo} and immersion of graphene in alcohol.~\cite{Xie2009cfo} Theoretical studies~\cite{Setton1996cnt, Tomanek2002mow, Braga2004sad, Pan2005ais, Chen2007sad, Mpourmpakis2007cna, Shi2009gbo, Patra2009naa} indicate that the lowest energy configuration of a large graphene flake is fully wrapped. However, it is separated from a flat state by a large energy barrier due to bending rigidity. This is one reason why flat free-standing graphene membranes can exist.~\cite{Booth2008mgm} In this paper we investigate another possible reason: the wrapping is arrested by unfavorable boundary conditions or external fields, and so what is observed is a partially scrolled edge of the otherwise mostly flat sample, see Fig.~\ref{fig:clamped}. We estimate dimensions of such edge CNSs, study the charge distribution therein, and discuss how they can be controlled electrically. Our approach relies on the continuum elasticity and classical electrostatics theories. It should be adequate for typical CNS, which contain anywhere from several tens to many thousands atoms around the circumference.

We begin with revisiting the problem of a CNS of a uniform cross-section with no axial elastic stress. The main structural property of interest is the smallest possible inner diameter $d_{\min}$ of a stable CNS. The input into our calculation are the following material parameters,~\cite{Kudin2001c2f, Tomanek2002mow, LWKH08, Spanu2009nas} which are derived from the properties of graphite and carbon nanotubes: the interlayer distance $h = 0.34\,\text{nm}$, the elastic bending modulus $D = 1.5\,\text{eV}$, the two-dimensional (2D) Young's modulus $Y_2 = 340\, {\text{N}}/{\text{m}} \approx 2100\, {\text{eV}}/{\text{nm}^2}$, and the van~der~Waals energy per unit area of contact $\gamma = (40 \pm 10)\,\text{meV}/\text{atom} = (1.5 \pm 0.4)\,\text{eV}/\text{nm}^2$.~\cite{Comment_on_vdW_1}

The scrolling is driven by the competition between van der Waals attraction of graphene to itself and its bending rigidity. For a large CNS the former dominates and the scrolling is advantageous. We find that the difference in energy $\varepsilon$ per unit length of a CNS and a flat sheet of the same area area has an approximately linear dependence
\begin{equation}
\varepsilon(s) \approx -(s - s_0) g\,,\quad s > s_0\,,
\label{eqn:E_fit}
\end{equation} 
on the total arc length $s$. In the practically important range $10 < s\, (\text{nm}) < 50$ we get $g = 1.2\,\text{eV}/\text{nm}$ and $s_0 = 12\,\text{nm}$. The coefficient $g$ is somewhat smaller than $\gamma$ because the bending energy is still significant at such $s$. At $s \gg s_0$, the derivative $\partial \varepsilon / \partial s$ tends to $-\gamma$.

\begin{figure}[t]
\begin{center}
\includegraphics[width=8.5cm]{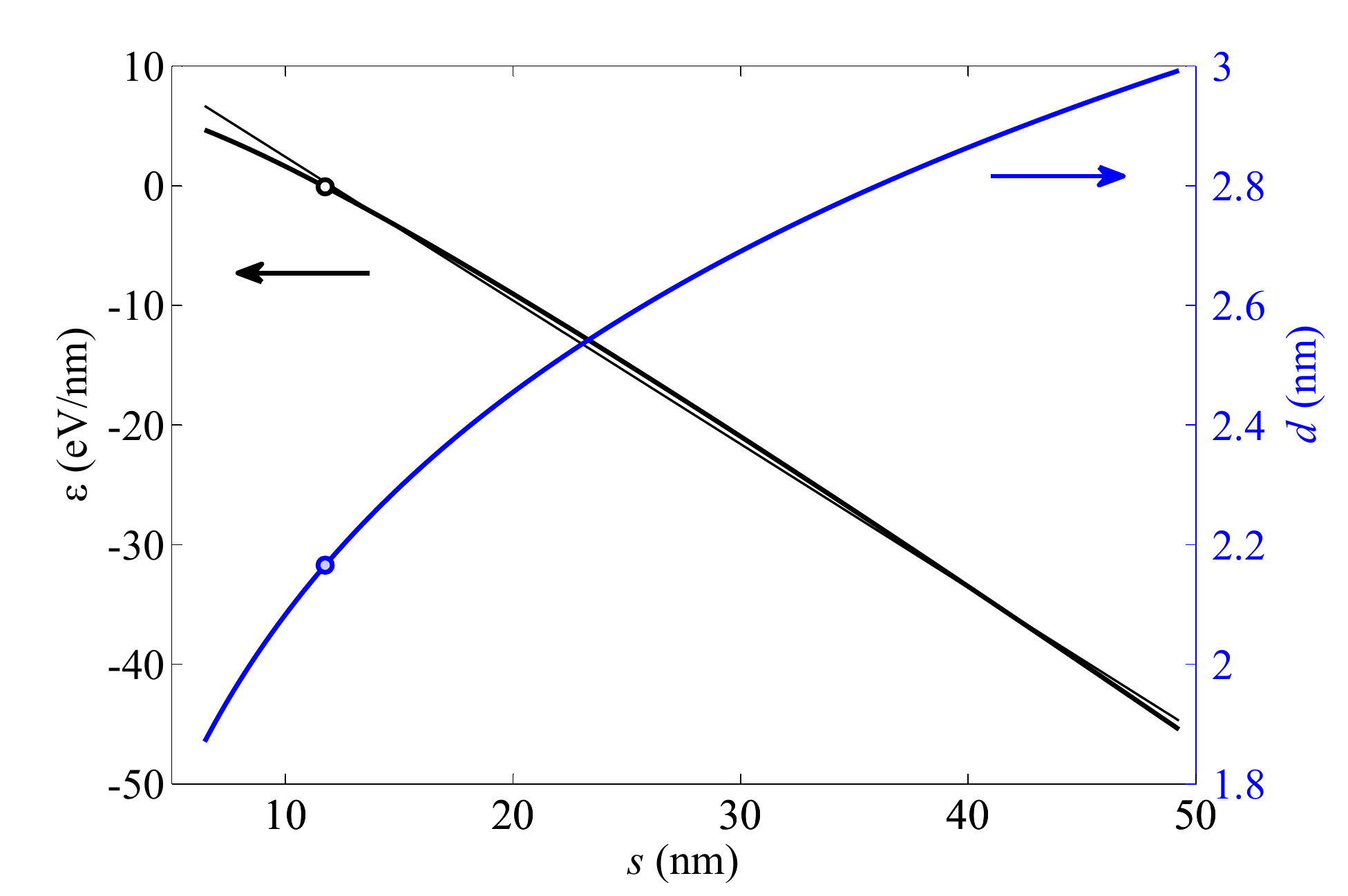}
\end{center}
\caption{(Color online) Energy and the inner diameter $d = 2 r + (h / 2)$ of a uniform CNS computed within the continuum elasticity model (thick lines). The markers correspond to the smallest stable CNS. The straight thin line is Eq.~\eqref{eqn:E_fit}.
\label{fig:uniscroll}
}
\end{figure}

To obtain these results we assume that the CNS is shaped as an Archimedian spiral~\cite{Setton1996cnt, Braga2004sad} $\rho(\phi) = a \phi$, where $\theta \leq \phi \leq \Theta$ is the polar angle in the cross-sectional plane and
$a \equiv h / (2 \pi)$. Ignoring any changes in the electron spectrum, we take $\varepsilon = \varepsilon_{vdW} + \varepsilon_{B}$ where $\varepsilon_B$ and $\varepsilon_{vdW}$ are the bending and the van~der~Waals energy, respectively. Expressing those in terms of the local curvature and the appropriate arc lengths, we get the energy functional $\varepsilon(\theta, \Theta)$ which can be easily minimized numerically. The results are shown in Fig.~\ref{fig:uniscroll}. Thus, the arc length $s_{\min} = 12\,\text{nm}$ of the smallest stable CNS is determined by $\varepsilon(s_{\min}) = 0$. The corresponding inner diameter is $d_{\min} = 2.2\,\text{nm}$, see Fig.~\ref{fig:uniscroll}. Our calculation avoids several approximations made in previous elasticity theory models of CNS~\cite{Tomanek2002mow, Shi2009gbo, Shi2009tcs} and as a result gives somewhat different $d_{\min}$ for the same input parameters. With the quoted values we achieve a nominal agreement~\cite{Comment_on_vdW_2} with molecular dynamics~\cite{Braga2004sad, Shi2009tcs} and \textit{ab initio} simulations.~\cite{Chen2007sad}

As seen in Fig.~\ref{fig:uniscroll}, there is a large energy barrier $\varepsilon(0) \sim 5\,\text{eV}/\text{nm}$ that separates the initial flat and the final curled states. This is why the optimal inner radius can hardly be achieved in practice.~\cite{Comment_on_Meyer} Realistic $r \gg d_{\min} / 2$ are presumably determined by interaction of graphene with scrolling ``chaperones,'' e.g., stray fluid particles.~\cite{Xie2009cfo, Patra2009naa} Similarly, the outer radius $R$ may never reach the maximum allowed by the total arc length $s$ if external conditions inhibit the scrolling. Below we consider two such conditions: mechanical clamping and electrostatic repulsion.


When a CNS forms at the free edge of an initially flat strip, it is often clamped at two other edges, see Fig.~\ref{fig:clamped}. In this geometry, finite in-plane elastic deformations build up as CNS rolls inward. Computing the corresponding equilibrium shape of the CNS is much more complicated than for the uniform scroll, and so we make further approximations. We assume that everywhere except near the corners, $s(x)$ and $R(x)$ are slowly varying functions of $|x| \leq L / 2$ that satisfy the inequalities $s(x) \gg R(x) \gg r \sim s_0$. This allows us to treat the CNS as a slender curved beam. Its axis stays predominantly in the plane and its shape $y(x)$ is dictated by the equilibrium between the force $-\partial \varepsilon / \partial s \simeq \gamma$ that pulls it in and the internal longitudinal tension $T$ that tries to keep it straight: $\gamma \simeq -T y^{\prime\prime}(x)$.
%
%
Thus, function $y(x)$ is parabolic:
\begin{equation}
y(x) \simeq y(0) \left(1 - \frac{4 x^2}{L^2} \right)\,,
\quad y(0) = \frac{\gamma L^2}{8 T}\,.
\label{eqn:y}
\end{equation}
We will now calculate $T$ and use it to show that
\begin{equation}
\frac{y(0)}{L} \simeq \left[
\frac{\gamma}{32 Y_2} \, \ln \left(\frac{L}{d_{\min}}\right)
\right]^{1/4} \approx 0.10\,.
\label{eqn:y_0}
\end{equation}
For example, if $L = 2\,\mu\text{m}$, then $y(0) \approx 20\,\text{nm}$, in agreement with observations, see Fig.~\ref{fig:clamped}.

The derivation of Eq.~\eqref{eqn:y_0} relies on the smallness of the ratio $\gamma / Y_2 \ll 1$. While the unscrolled portion of graphene is stretched outward by the same force per unit length $\gamma$, the characteristic elastic deformation is $u(x) \sim (\gamma / Y_2) L (1 - 4 x^2 / L^2)^{1 - \alpha}$. Here the exponent~\cite{Prada2010ses} $\alpha \approx 0.16$ accounts for the power-law singularities near the corners.~\cite{Williams1952ssr} From Eq.~\eqref{eqn:y_0}, we see that $u(x)$ is parametrically smaller than $y(x)$, and so the unwrapped portion of graphene acts as an inextensible membrane. This implies that the ``unstretched'' ($T = 0$) shape of the beam is the parabolic arc located halfway between $y = 0$ and the actual beam position. The tension $T$ is related to the length difference $\Delta L$ of these two parabolas:
\begin{equation}
\Delta L \simeq \frac38
                \int\limits_{-L / 2}^{L / 2} d x [y^\prime(x)]^2
              = \frac{2}{L}\, y^2(0) \simeq \frac{T}{Y_2}
                  \int\limits_{-L / 2}^{L / 2} \frac{d x}{y(x)}\,.
\label{eqn:DeltaL}
\end{equation}
The denominator $y(x)$ in the last integral takes into account that the local cross-sectional area of the CNS is $s(x) h$ with $s(x) \simeq y(x)$ for the inextensible graphene. The integral logarithmically diverges at the end points where our approximation is invalid. Imposing a short-distance cutoff $d_{\min}$, we recover Eq.~\eqref{eqn:y_0}. (The numerical answer given is nearly independent of this cutoff.) Finally, the radius $R(x)$ of the CNS can be obtained from $s(x)$:
$
  R(x) \simeq \sqrt{2 a s(x)} \simeq \sqrt{2 a y(0) (1 - 4 x^2 / L^2)}
$.

\begin{figure}[t]
\begin{center}
\includegraphics[width=8cm]{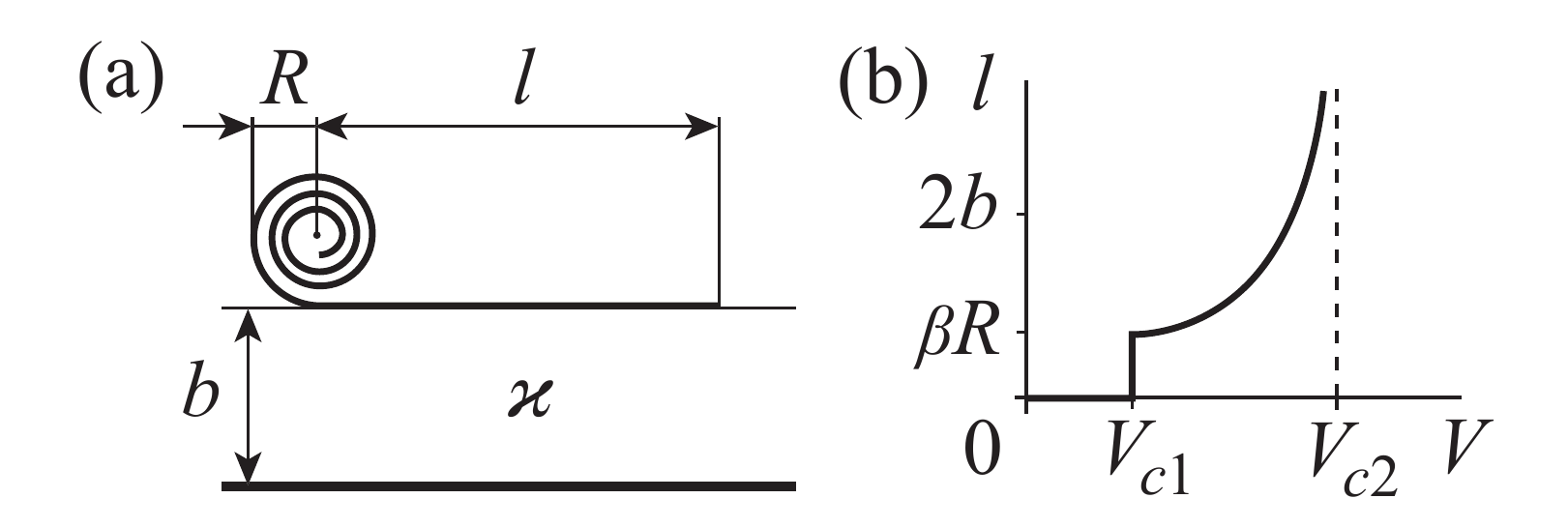}
\end{center}
\caption{(a) Model geometry (see text) (b) Optimal $l$ \textit{vs}. $V$.
\label{fig:eyelid}
}
\end{figure}

Let us now go back to the case of a uniform scroll but add interaction with substrate and external electrodes. Suppose the sample is deposited on an insulating substrate of thickness $b$ and dielectric constant $\varkappa$. The other side of the substrate is covered with a metallic gate, see Fig.~\ref{fig:eyelid}(a). If a voltage $V$ is applied between the graphene and the gate, the total energy of the system acquires the electrostatic contribution $\varepsilon_{el} = C V^2 / 2 - q V = -C V^2 / 2$, where $C$ and $q = C V$ are the capacitance and charge per unit length of graphene, respectively. (The first term is the charging energy, the second one is the work of external sources). Suppose that the CNS can form only at one edge and let $l$ be the width of the unwrapped portion. Taking the fully wrapped state, $l = 0$, to be the energy reference point, we get
\begin{equation}
\varepsilon(l) = g l - [C(l) - C(0)] V^2 / 2\,.
\label{eqn:e}
\end{equation}
Since $C(l)$ increases with $l$, the charging of graphene would tend to unwrap the CNS. We will show that there is a critical voltage
\begin{equation}
V_{c2} = \sqrt{{8 \pi} g b / {\varkappa}}\,,
\label{eqn:V_c2}
\end{equation}
above which the unwrapping is complete, $l = s$. Here we assume that the CNS is thick enough, $R \gg \sqrt{D / g}$, so that the effect of the bending modulus $D$ on $g$ can be neglected. However, $g = \gamma - \gamma_s$ can be reduced from $\gamma$ due to the van der Waals coupling $\gamma_s$ to the substrate. 

For a quick estimate, consider the standard SiO$_2$ substrate ($\varkappa = 3.9$, $b = 300\,\text{nm}$). Unfortunately, reported $\gamma_s$ range widely, from about one half~\cite{Hertel1998doc, Betal08b} (similar to metal substrates~\cite{Vanin2010gom}) to one hundredth~\cite{Sabio2008eib, Stolyarova2009oog} of $\gamma$ itself. Presumably, $\gamma_s$ is strongly influenced by surface conditions. Assuming $\gamma_s \ll \gamma$, we get the upper bound on the critical voltage, $V_{c2} = 65\,\text{V}$, which is still in the experimental range. The corresponding carrier density is $n_{c2} = \varkappa V_{c2} / (4 \pi e b) = 4.6\times 10^{12}\,\text{cm}^2$.

In order to derive Eq.~\eqref{eqn:V_c2} and find the optimal $l$ at $V < V_{c2}$ we need to have a model for the capacitance $C(V)$. It is well known that $C$ involves both classical and quantum terms. The former is the capacitance computed treating the CNS as a metallic cylinder whose charge resides only on its surface. Quantum effects make the charge to spread to a depth of the order of the screening length, reducing the capacitance. This length can be estimated using the Thomas-Fermi (TF) approximation. Locally, the CNS is similar to a stack of 2D layers with the total density of states $\nu_1 / h$ where $\nu_1$ is the density of states of a single layer. The TF screening length of such a stack is
$
r_{\text{TF}} = \sqrt{h / (4 \pi \nu_1)} = \sqrt{r_{\parallel} h / 2}
$,
where $r_{\parallel} = (2 \pi e^2 \nu_1)^{-1}$ is the in-plane screening length (for refinements, see Ref.~\onlinecite{Visscher1971dsi}). For $n \sim 10^{12}\,\text{cm}^{-2}$ we have $r_{\parallel} \sim 10\,\text{nm}$, so that $r_{\text{TF}} \sim 1\,\text{nm} \approx 3h$. Thus, for CNS of four or more coils, the classical capacitance is a good first approximation. Similarly, if $l \gg r_{\parallel}$, the unwrapped portion of graphene can be treated as a metallic strip.

Assuming both such conditions are met, we can use some standard expressions for $C(l)$. If $l \gg 2 b \gg R$, then $C(l)$ is dominated by the flat piece and is given by the parallel-plate capacitor formula
\begin{equation}
C(l) = \frac{\varkappa}{2\pi} \left[\frac{l}{2b} + t_b(\varkappa)\right] + \frac{1}{2\pi^2} \left[\ln \left(\frac{l}{2 b}\right) + t_t(\varkappa)\right]
\label{eqn:C_large_l}
\end{equation}
(exact values of $t_b, t_t \sim 1$
%
%
are not important here), in which case Eq.~\eqref{eqn:e} takes the form
\begin{equation}
\label{eqn:e_large_l}
\varepsilon = \tilde{\gamma} l
            - \frac{V^2}{4\pi^2} \ln \left(\frac{l}{2 b}\right)
            + \text{const}\,,
\quad
\tilde{\gamma} = g - \frac{\varkappa V^2}{8\pi b}\,.
\end{equation}
In complete analogy to electrowetting,~\cite{Mugele2005efb} the applied voltage reduces the effective surface tension $\tilde\gamma$. The critical voltage $V_{c2}$ [Eq.~\eqref{eqn:V_c2}] is determined simply by $\tilde\gamma(V_{c2}) = 0$. The origin of the force that unwraps the CNS is the asymmetry of the fringing field of the capacitor. This field is stronger on the outward side of the CNS because there it is not screened by the flat piece of graphene. Hence, a net torque on the CNS is produced.

At $V < V_{c2}$ the surface tension $\tilde\gamma$ is positive and the scrolling becomes possible. Minimizing $\varepsilon(l)$, we obtain the optimal $l$ as follows:
\begin{equation}
l = \frac{2}{\pi \varkappa}\, \frac{b}{(V_{c2} / V)^2 - 1}\,.
\label{eqn:l_I}
\end{equation}
For $\varkappa = 3.9$, the range of validity of this formula is narrow,
$0.95 < V / V_{c2} < 1$, because at lower $V$ we have $l < 2b$. Nevertheless, it does indicate that if we start with a flat sample of say micron dimensions at $V = V_{c2}$ and reduce $V$ by just $5\%$, a rapid growth of the CNS at the expense of the unwrapped portion would occur.

At still lower voltages, $V \lesssim 0.8 V_{c2}$, at which $l$ should be several times smaller than $2 b$, we can use another analytic approximation:
\begin{equation}
C(l) = \frac{\varkappa + 1}{4} \frac{1}{\ln \big(8 c(\varkappa) b / l_{\text{eff}} \big)}\,,
\label{eqn:C_small_l}
\end{equation}
where $l_{\text{eff}}$ is the effective width of the strip plus CNS:
\begin{align}
l_{\text{eff}} &\simeq l + \frac{2 \pi}{3} R + \frac{4 \pi^2}{27}\, \frac{R^2}{l}\,,
 &\text{if }l \gg R\,,\\
 &\simeq 4 R + \frac{1}{3\pi}\, \frac{l^3}{R^2}\,,
 &\text{if }l \ll R\,.
\label{eqn:l_eff}
\end{align}
We derived these equations using the standard technique of conformal mapping followed by the series expansion to the order indicated. In the limiting cases $R = 0$ and $l = 0$, they reproduce the known results for the metallic strip~\cite{Silvestrov2008caa} and the metallic cylinder, respectively. Function $c(\varkappa)$ in Eq.~\eqref{eqn:C_small_l} is given by a certain cumbersome integral, which is however bounded and smoothly varying. Its characteristic values are $c(1) = 1$, $c(3.9) = \exp(-0.31) = 0.73$, and $c(\infty) = 2 / \pi = 0.64$.

Analysis of the energy functional~\eqref{eqn:e} using Eqs.~\eqref{eqn:C_small_l}--\eqref{eqn:l_eff} unravels a new feature. There exists a lower critical voltage $V_{c1}$,
\begin{equation}
\frac{V_{c1}}{V_{c2}} \simeq \ln \left[\frac{2 c(\varkappa) b}{\beta R} \right]
\sqrt{\frac{\beta / \pi}{1 + \varkappa^{-1} }\, \frac{R}{b}}\,,
\quad \beta = 5.4\,,
\label{eqn:V_c1}
\end{equation}
at which the optimal width $l$ discontinuously jumps from $\beta R$ to zero, i.e., the sheet becomes fully wrapped. For $R = 10$--$20\,\text{nm}$ and $b = 300\,\text{nm}$, we get $V_{c1} \approx 0.4 V_{c2}$. Note that the $l = 0$ state is a local energy minimum at any $V$. This is because at $l \ll R$ the field is symmetric on both sides of the CNS and no net torque exists to promote its unwrapping. This may also be the reason why no unwrapping has been seen in prior simulations~\cite{Braga2004sad} even at carrier densities $n \sim 10^{14}\,\text{cm}^2$, i.e., hundred times higher than our $n_{c2}$. This leads us to believe that the transition to and from the fully wrapped CNS state would be hysteretic in practice. Of course, no bistability would exist if the strip widths $l \leq \beta R$ are unattainable due to clamping or obstacles on the CNS path. Finally, Eq.~\eqref{eqn:V_c1} could be modified if the CNS length $L$ varies as it wraps along, because of a non-rectangular sample shape.



In closing, we posit that electrostatic and strain-based control over the shape and size of CNSs promise new applications, such as ``eyelid'' actuators~\cite{Goodwin-Johansson2002rva} or valves in lab-on-a-chip devices. Being ultralight, they would operate in a THz range.~\cite{Braga2004sad, Shi2009tcs, Patra2009naa, Shi2010atn} Electronic paper and metamaterials may be created with CNSs made of optically active multilayer graphene. Electronic applications may include varactors and THz oscillators.~\cite{Shi2009tcs}

Recently two other publications appeared~\cite{Shi2010atn, Shi2010twc} where the effect of longitudinal electric field on CNS was considered. Similar to our case (the transverse field) it should create an accumulation of charge carriers, except they would have opposite polarity at the two ends of the CNS. The Coulomb repulsion of these mobile charges is an important energy correction, which can be computed by a method similar to ours. In Ref.~\onlinecite{Shi2010twc} only the polarization of immobile core states was included. We thank N.~Pugno for drawing our attention to these papers.


MMF and AHCN thank the Aspen Center for Physics for hospitality at the early stage of this work. MMF is supported by the NSF Grant DMR-0706654. AHCN acknowledges support from DOE Grant DE-FG02-08ER46512 and ONR Grant MURI N00014-09-1-1063. FG is supported by MEC (Spain) Grants FIS2008-00124 and CONSOLIDER CSD2007-00010, and also the Comunidad de Madrid, through CITECNOMIK.
We thank A.~Geim for useful discussions and providing the image shown in Fig~\ref{fig:clamped}.

%
\end{document}